\documentclass[dvips,twocolumn,APS,showpacs,amssymb,10pt]{revtex4}
\usepackage{graphicx}
\usepackage{bm}
\usepackage{epsfig}
\usepackage{color}
\usepackage{ulem}
\usepackage{verbatim}
\usepackage{amsmath}
\usepackage{amssymb}
\def\ketm#1{  \left\vert  #1   \right\rangle   }
\def\ket#1{  \left\vert  #1   \right\rangle   }

\def\bram#1{  \left\langle  #1   \right\vert   }

\def\etal{et al.}                                  
\newcommand{\bl}[1]{{\color{blue} #1}}

\renewcommand{\vec}[1]{\bm{#1}}
\newcommand{\ds}{\displaystyle}
\newcommand{\braa}[1]{\mathinner{\langle{#1}|}}
\newcommand{\kett}[1]{\mathinner{|{#1}\rangle}}
\begin{document}
\preprint{}
%
\title{Double-Electromagnetically Induced Transparency in a Y-type atomic system}
%
%
\author{L. Safari$^{1,}$\footnote{laleh.safari@oulu.fi}, D. Iablonskyi$^{1}$, and F. Fratini$^{1,2,3}$}
\affiliation{\it
$^1$ Department of Physics, University of Oulu, Box 3000, FI-90014 Oulu, Finland \\
$^2$ Departamento de F\'isica, Universidade Federal de Minas Gerais, 30123-970 Belo Horizonte, Brazil\\
$^3$ Institut N\'eel-CNRS, BP 166, 25 rue des Martyrs, 38042 Grenoble Cedex 9, France}
\date{\today \\[0.3cm]}%
%
%
%
\begin{abstract}
We study the absorption and dispersion properties of a weak tunable probe field in a four-level Y-type atomic system driven by two strong laser (coupling) fields within the framework of density matrix formalism. It is found that the probe absorption profile displays double-electromagnetically induced transparency (double-EIT) and it is shown how to control it by changing the Rabi frequencies as well as the atom field detuning of the coupling fields. 
\end{abstract}	
\maketitle
%
%
%
%
%
%
\section{Introduction}\label{sec:intoduc}

Atomic coherence and interference are two fundamental phenomena which can modify the spectral properties of a multilevel atomic system. They lead to novel phenomena in quantum optics such as
electromagnetically induced transparency \cite{Fleischhauer:05,Harris:97,S.N.Nikolic:13}, electromagnetically induced absorption (EIA) \cite{Evers:10,Lipsich:00}, coherent population trapping (CPT) \cite{Stähler:02}, lasing without inversion (LWI) \cite{Zhu:96,Kozlov:06,Koch1986}, inversion without lasing \cite{Kozlov:06, MarlanO.scully:89}, quenching of spontaneous emission \cite{Kishore:03}, superluminal light propagation \cite{L.J.Wang:00,A.Kuzmich:01,mousavi:10,M.Mahmoudi:2009,Harris:95} and other phenomena.

Generally, quantum interference takes place when there are two or more indistinguishable pathways for a transition \cite{Scully/Zubairy:97, L.Safari:12A, L.Safari:12B}. EIT is a quantum interference phenomenon that can make a normally opaque medium completely transparent for a probe beam. Such a transparency is due to the destructive interference between the probe absorption amplitudes. Since its first experimental demonstration in 1991  by Boller \textit{et al.} \cite{Boller:91}, EIT has been a subject of enhanced interest in scientific research. One of the remarkable effects EIT is capable of is to slow the propagation of light down up to hundreds of times \cite{A.M.Akulshin:2003,Hau:Harris:99,BinWu:10,Matthew:03} or even to almost stop light \cite{Bajcsy:03}.
For instance, the group velocity of a light pulse can be reduced to 17 meters per second in Bose-Einstein condensate of sodium atom gas \cite{Hau:Harris:99}. EIT in simple three-level $\Lambda$, $V$ and cascade (ladder) atomic systems has been widely investigated in the literature \cite{Harris:97,Hou.Wang:04,S.Mitra:13,M.Xiao:1995}. In contrast there have been few theoretical and experimental studies on the four-level Y-type atomic system \cite{Hou:04,Dutta:08,Gao:00,A.B.Mirza:2012}. For example, Hou \textit{et al.} \cite{Hou:04} investigated the effect of vacuum-induced coherence (VIC) on absorption properties in Y-type atomic system. Dutta \textit{et al.} \cite{Dutta:08} studied the VIC effect on absorption properties of Y-type four-level system by applying incoherent pump fields. Moreover, Gao \textit{et al.} \cite{Gao:00} observed electromagnetically induced inhibition of two-photon absorption in sodium vapor.

As known from three-level atomic system \cite{Fleischhauer:05}, the action of one strong coupling field gives rise to an EIT window in the absorption profile of the probe field and to subluminal light propagation. In contrast to the three-level atomic system, however, a second EIT window can occur and can be modified in a four-level atomic system due to the additional (second) coupling field. Such a double-EIT system has been subject of a few theoretical and experimental studies since it may enhance nonlinear optical processes in comparison with single EIT systems \cite{Y.Zhang:07,Z.Y.Zhao:12,Y.Liu:12,Z.Chang:12,L-G.Si:10}. In this work, we consider a four-level Y-type atomic system which is driven by a weak and tunable probe field as well as by two coherent coupling fields. For such a system, we analyze how the dispersion and absorption of the probe field are affected by the Rabi frequencies as well as by the detuning of the coupling fields. We show that two EIT windows appear and that their occurrence can be also understood within the \textit{dressed} state picture of the upper levels and their transitions to and from the ground level.

The work is organized as follows. In subsection \ref{subsec:A}, we first introduce our Y-type atomic system and various laser fields. We then derive the density matrix equations of motion for this system (subsection \ref{subsec:B}). In subsection \ref{subsec:C}, we explain how the steady-state solution is related to the linear susceptibility of the atomic system and the group velocity of the probe field. In subsection \ref{subsec:D}, we derive an analytical expression for the linear susceptibility based on the analytical steady state solution. In subsection \ref{subsec:E}, the dressed states and their energies are obtained by diagonalizing the hamiltonian of the atom plus the coupling fields. In section \ref{Sec:III}, the absorption and dispersion of the probe field are investigated for several cases and analyzed by means of the dressed state picture. Finally, a few conclusions are given in section \ref{sec:conclusion}.
%
%
%
%
%
%
\section{Theoretical formulation} \label{sec.basics}
%
%
\subsection{The four-level Y-type atomic system}
\label{subsec:A}
\begin{figure}[t]
\includegraphics[scale=0.6]{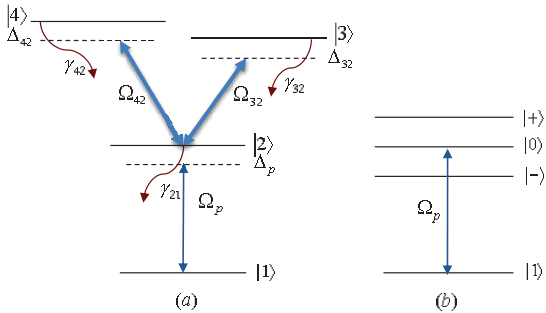}
\caption{Four-level Y-type driven atom and its $dressed$ analog. (a) Levels and energy scheme in the {\it bare} basis set. (b) Levels and energy scheme in the {\it dressed} basis set. The symbols $\Delta_{ij}$, $\Omega_{ij}$ and $\gamma_{ij}$ are defined in the text.}
\label{fig:four-level-system}
\end{figure}

Let us start with introducing the four-level Y-type model system which is considered in this work and displayed in figure~\ref{fig:four-level-system}(a). This model contains the ground level $\ketm{1}$, an intermediate level $\ketm{2}$ and the two upper levels $\ketm{3}$ and $\ketm{4}$ which might be close to each other in energy or even (nearly) degenerate. The levels $\ketm{1}$ and $\ketm{2}$ are coupled by the weak and tunable probe-field laser with Rabi frequency $\Omega_{p}$, while the intermediate level $\ketm{2}$ is coupled also to $\ketm{3}$ and $\ketm{4}$ by two distinguishable tunable coupling fields with Rabi frequencies $\Omega_{32}$ and $\Omega_{42}$, respectively. In this notation, the Rabi frequencies 
$\Omega_{k2} \,=\, \mathbf{d}_{k2}\,\cdot\,\mathbf{E}_{k2} / \hbar$, with $ k\,=\, 3,4$, describe the strength of the two coupling fields and are usually considered to be much larger than the Rabi frequency of the probe field, $\Omega_{p}\,=\, \textbf{d}_{21}\,\cdot\,\textbf{E}_p / \hbar$. Here, $\mathbf{d}_{ik}$ denote the dipole moments of the atomic transitions (which are properties of the atom) and $\mathbf{E}_{ik}$ the corresponding electric field amplitudes (which are properties of the associated laser fields), while $\hbar$ is reduced Planck's constant. Apart from the ground level $\ketm{1}$, all levels decay via spontaneous emission of photons, and with given decay rates $\gamma_{21},\; \gamma_{32}$ and $\gamma_{42}$. As usual, we suppose that the upper levels $\ketm{3}$ and $\ketm{4}$ decay only to the intermediate level $\ketm{2}$ owing to their parities and total angular momenta. Such a  Y-type model system is approximately realized, for example, in rubidium vapor if we identify the $(4p^6 5s)\; ^2S_{1/2}$ ground level of neutral rubidium with level $\ketm{1}$, the $(4p^6 5p) \; ^2P_{3/2}$ with level  $\ketm{2}$ and the two 
$(4p^6 5d)\; ^2D_{3/2,5/2}$ levels with the upper levels $\ketm{3}$ and $\ketm{4}$. For this assignment of the given levels, the three transitions $5s_{1/2}\,\leftrightarrow\, 5p_{3/2}$, $5p_{3/2}\,\leftrightarrow\, 4d_{5/2}$ and $5p_{3/2}\,\leftrightarrow\, 4d_{3/2}$ are (electric-dipole) \textit{allowed}, while the direct transition of levels $\ketm{3}$ and $\ketm{4}$ to the ground level are dipole-forbidden. Below, we shall denote the laser field detunings, which are the difference between atomic transition frequencies and laser field frequencies, by $\Delta_{32}=\omega_{32}-\omega_{c1}$, $\Delta_{42} = \omega_{42}-\omega_{c2}$ and $\Delta_{p} = \omega_{21}-\omega_{p}$ where $\omega_{c1}$, $\omega_{c2}$ and $\omega_{p}$ refer to the frequencies of the two coupling and the probe fields, respectively. Atomic transition frequencies are denoted by $\omega_{ik} = (E_i-E_k) / \hbar$ for the transition $\ketm{i} \leftrightarrow \ketm{k}$.

In the interaction picture, the total Hamiltonian of the four-level Y-type system can be written as
\begin{eqnarray}
\label{totalH}
   H^{int} & = & H_{0} \,+\, H^{int}_{I}
\end{eqnarray}
with the atomic (field-free) Hamiltonian
\begin{eqnarray}
\label{Free-H}        
   H_{0} & = & \hbar\omega_1\,\ketm{1}\bram{1} \,+\,
               \hbar\omega_2\,\ketm{2}\bram{2} 
   \nonumber \\[0.1cm] 
         & + & \hbar\omega_3\,\ketm{3}\bram{3} \,+\,
               \hbar\omega_4\,\ketm{4}\bram{4} 
\end{eqnarray}
and the interaction Hamiltonian
\begin{eqnarray}
\label{Interaction-H}
   H^{int}_I & = &
   -\, \hbar\Omega_{p}\,  e^{-i\Delta_p    t} \ketm{2}\bram{1} \,-\,
       \hbar\Omega_{32}\, e^{-i\Delta_{32} t} \ketm{3}\bram{2} \,
   \nonumber \\[0.1cm]
   &  & 
   -\, \hbar\Omega_{42}\, e^{-i\Delta_{42} t} \ketm{4}\bram{2} \,+\,
   \text{h.c.} \, ,
\end{eqnarray}
and where, as usual, we made use of the rotating wave and dipole approximations. 
As seen from equation (\ref{Interaction-H}), the interaction Hamiltonian just describes the excitation and de-excitation of the system due to the three employed laser fields.
%
%
%
\subsection{Density matrix equation of motion}
\label{subsec:B}
Having the Hamiltonian of the four-level Y-type system, \eqref{totalH}, we can apply Liouville's equation \cite{Scully/Zubairy:97, mousavi:10}
\begin{eqnarray}
\label{liuvil-equ}
   \dot{\rho} & = & -i/\hbar \, [H_I^{int},\rho] \:+\: L\,\rho \, ,
\end{eqnarray} 
in order to determine the time-evolution of its density matrix, and where $L\,\rho$ represents the relaxation of the system due to its (effective) interaction with the environment \cite{Scully/Zubairy:97}. For the given system, equation~(\ref{liuvil-equ}) represents a coupled set of nine independent, first-order differential equations due to the hermiticity and normalization of the density matrix. It can be solved, in principle, for any proper initial condition  $\rho(t_0)$ in order to obtain the population (diagonal matrix elements) and coherences (off-diagonal elements) of the four-level system. However, before we solve this set of equations, we shall remove the explicit time-dependent factors in the interaction Hamiltonian by moving to a \textit{rotating} frame, i.e.\ by absorbing these factors into 
the nondiagonal matrix elements. With these considerations, Eq. (\ref{liuvil-equ}) can be cast into the form
\begin{widetext}
\begin{subequations}
\begin{align}
		\label{density-matrix_a}
   \dot{\rho}_{11} &= 2\gamma_{21}\,\rho_{22}   \,+\, 
                       (i\Omega_{p}\,\rho_{21}   \,+\, h.c.)
												~, 
   \displaybreak[0]\\[0.1cm] 
	   \dot{\rho}_{33} &= -2\gamma_{32}\,\rho_{33}  \,-\,
                        (i\Omega_{32}\,\rho_{32} \,+\, h.c.)\,
												-\,\eta\,(\rho_{43}^*+\rho_{43})~,
   \displaybreak[0]\\[0.1cm]                    
   \dot{\rho}_{44} &= -2\gamma_{42}\,\rho_{44}  \,-\,
                        (i\Omega_{42}\,\rho_{42} \,+\, h.c.)\,
												-\,\eta\,(\rho_{43}^*+\rho_{43})~,
   \displaybreak[0]\\[0.1cm]
   \dot{\rho}_{13} &= -(\gamma_{32}-i\Delta_{p}-i\Delta_{32})\,\rho_{13} \,-\,
                        i\Omega_{32}\,\rho_{12} \,+\,
                        i\Omega_{p}\,\rho_{23} \,
												-\,\eta\,\rho_{41}^*~,
   \displaybreak[0]\\[0.1cm]
   \dot{\rho}_{21} &= -(\gamma_{21}+i\Delta_{p})\,\rho_{21} \,+\,
                        i\Omega_{42}\rho_{41}
												-i\Omega_{p}^{\ast}\,\rho_{22} \,+\,
                        i\Omega_{p}^{\ast}\,\rho_{11} \,+\,
                        i\Omega_{32}\,\rho_{13}^*~,          
   \displaybreak[0]\\[0.1cm]
   \dot{\rho}_{23} &= -(\gamma_{32}+\gamma_{21}-i\Delta_{32})\,\rho_{23} \,+\,
                        i\Omega_{42}\rho_{43}-i\Omega_{32}\,\rho_{22} \,+\,
                        i\Omega_{p}^{\ast}\rho_{13}+i\Omega_{32}\,\rho_{33}\,
												-\,\eta\,\rho_{42}^*~, 
   \displaybreak[0]\\[0.1cm]
   \dot{\rho}_{41} &= -(\gamma_{42}+i\Delta_{42}+i\Delta_{p})\,\rho_{41} \,-\,
                        i\Omega_{p}^{\ast}\,\rho_{42} \,+\,
                        i\Omega_{42}^{\ast}\,\rho_{21}\,
												-\,\eta\,\rho_{13}^*~,   
  \displaybreak[0] \\[0.1cm]
   \dot{\rho}_{42} &= -(\gamma_{42}+\gamma_{21}+i\Delta_{42})\,\rho_{42} \,-\,
                        i\Omega_{42}^{\ast}\,\rho_{44} \,-\,
                        i\Omega_{p}\, \rho_{41} \,-\,
                        i\Omega_{32}^{\ast}\,\rho_{43} \,+\,
                        i\Omega_{42}^{\ast}\,\rho_{22}\,
												-\,\eta\,\rho_{23}^*~,
  \displaybreak[0] \\[0.1cm]
	\dot{\rho}_{43} &=  -(\gamma_{42}+\gamma_{32} +i\Delta_{42}-i\Delta_{32}) 
                        \, \rho_{43} \,-\,
                        i\Omega_{32}\,\rho_{42} \,+\,
                        i\Omega_{42}^{\ast}\,\rho_{23}\,
												-\,\eta\,(\rho_{33}+\rho_{44})~,
 \displaybreak[0] \\[0.1cm]
 \label{density-matrix_i}
 &\hspace{-0.5cm}\rho_{11}+\rho_{22}+\rho_{33}+\rho_{44}= 1
\end{align}
\end{subequations}
\end{widetext}
where $\gamma_{ij}$ originate from the relaxation term in equation \eqref{liuvil-equ}, and $\eta=p\sqrt{\gamma_{32}\gamma_{42}}$ represents the VIC effects resulted from the cross-coupling between two decay paths $\kett{4}\to\kett{2}$ and $\kett{3}\to\kett{2}$ \cite{Hou.Wang:04}. The parameter $p$ is defined as $p=\textrm{\bf d}_{32}\cdot\textrm{\bf d}_{42}/(|\textrm{\bf d}_{32}|\,|\textrm{\bf d}_{42}|)$, and thus ranges from $0$ to $1$. In the following, we shall set $p=0$, i.e. we shall suppose the dipole $\textrm{\bf d}_{32}$ and $\textrm{\bf d}_{42}$ be orthogonal with each other. Different settings for $p$ do not qualitatively change the absorption and dispersion peaks: The position of the peaks and the sign of the first derivative of the peaks' slopes are not affected. Rather, larger values for $p$ only modify the intensity and the width of the absorption and dispersion peaks. Moreover, it is found that VIC effects are important only for Rabi frequencies of the order of or bigger than the decay rates, viz. only for $\Omega_{32}\gtrsim \gamma_{32}$, $\Omega_{42}\gtrsim \gamma_{42}$.

Below, we are mainly interested in the coherency between the levels $\ketm{1}$ and $\ketm{2}$, which is represented by the element $\rho_{21}$. On the one hand, by supposing both the probe and the coupling fields to be continuous-wave lasers irradiating the system and by setting $\dot{\rho}_{ij} = 0$, Eqs.~(\ref{density-matrix_a}-\ref{density-matrix_i}) can be analytically solved  and the solutions are called \textit{steady} state solutions. We will follow this method in Subsec. \ref{subsec:D}, within the weak-probe approximation. On the other hand, such a steady state solutions can be also obtained numerically by directly solving  Eqs.~(\ref{density-matrix_a}-\ref{density-matrix_i}) for a sufficient long time with the condition that the atom is initially (at $t=0$, before the fields are switched on) in its ground level $\ketm{1}$, i.e. $\rho_{11}(0) = 1$ and $\rho_{ij}(0) = 0$ otherwise. In Sec. \ref{Sec:III}, we checked that both solutions lead to the same results for absorption and dispersion of the probe field. 
%
%
%
\subsection{Susceptibility and group velocity}
\label{subsec:C}

In most EIT experiments, the electrical susceptibility of the atomic target is measured close to some resonant frequencies of the atoms in order to determine the polarizability of the medium as (linear) response to the applied laser fields. Though the susceptibility is a macroscopic property of the medium, it is also related to the coherency $\rho_{21}$ and the dipole moment of the underlying probe transition by the relation \cite{Scully/Zubairy:97}
\begin{eqnarray}
\label{susceptibility}
   \chi & = & \frac{N\,|\vec{d}_{21}|^2}{\epsilon_0\,\hbar}\: \rho_{21} \, ,
\end{eqnarray} 
where $N$ is the number density of atoms in the medium and $\epsilon_0$ the vacuum permittivity. In particular, the real and imaginary parts of the susceptibility, $\chi=\chi'+i\chi''$, refer to the dispersion and absorption of the probe field by the medium. We can, moreover, utilize the linear susceptibility in order to express the probe group velocity \cite{Agarwal2001}
\begin{eqnarray}
\label{group-velocity}
   v_g & = & \frac{c}{1 \,+\, 2\pi\,\chi'(\omega_p) \,+\,
                      2\pi\, \omega_p\,
                      \frac{\partial}{\partial\omega_p}\,\chi'(\omega_p)}~,
\end{eqnarray} 
where $c$ denotes the speed of light in vacuum. As seen from this expression, both the dispersion $\chi'$ as well as its derivative with regard to the frequency of the probe field, $\frac{\partial}{\partial\omega_p}\chi'(\omega_p)$, affect the propagation of the probe field through the medium. Apart from a small dispersion $\chi'$ of the probe field, therefore, a quite sizable change in the group velocity can occur if its derivative $\frac{\partial}{\partial\omega_p}\chi'(\omega_p)$ becomes sufficiently large. Let us remark here that, due to the relation  $\partial \Delta_{p} = - \partial \omega_p$, a positive derivative of the dispersion with respect to $\Delta_p$ refers to a group velocity larger than $c$ (superluminal light propagation) while a negative derivative to a group velocity smaller than $c$ (subluminal light propagation). Moreover, a positive (negative) value of $\chi''$ = Im($\chi$) refers to an absorption (amplification) of the probe field.
%
%
%
\subsection{Analytical steady state solutions}
\label{subsec:D}

Relation (\ref{susceptibility}) is here utilized, together with the analytical solution for $\rho_{21}$, in order to express the dispersion $\chi'$ and absorption $\chi''$ of the probe field in terms of the probe detuning $\Delta_p$. Since the strength  of the probe field is supposed to be weak, when compared to the coupling fields ($\Omega_p << \Omega_{32}, \, \Omega_{42}$), we may just keep the terms \textit{linear} in the probe field, while the (Rabi frequencies of the) coupling fields are kept to all orders. With these assumptions, the relevant density matrix equations are found to be \cite{M.Mahmoudi:2009}
\begin{subequations}
\begin{eqnarray}
\label{important-equ}
   \dot{\rho}_{31} &=&-(\gamma_{32}+i\Delta_{p}+i\Delta_{32})\rho_{31}
                       +i\Omega_{32}^{\ast}\rho_{21}~,   \\
   \dot{\rho}_{21} &=&-(\gamma_{21}+i\Delta_{p})\rho_{21}
                       +i\Omega_{42}\rho_{41}-i\Omega_{p}^{\ast} \nonumber \\
                     & &+\,i\Omega_{32}\rho_{31}~,  \\
   \dot{\rho}_{41} &=&+(\gamma_{42}+i\Delta_{42}+i\Delta_{p})\rho_{41}  
                        +i\Omega_{42} ^{\ast}\rho_{21}~.
\end{eqnarray}
\end{subequations}
The steady state solution for $\rho_{21}$ can be derived by setting $\dot{\rho}_{ij} = 0$ in the above equations as
\begin{eqnarray}
\label{probe coherency}
&&\rho_{21}= \nonumber \\ 
&&\frac{-i\Omega^*_p (\gamma_{42}+i\Delta_{42}   
      +i\Delta_{p})(\gamma_{32}+i\Delta_{32}+i\Delta_{p})}{A+iB}
\end{eqnarray}
with
\begin{eqnarray}
\label{R}
           A &=& -\Delta_p\Big(\Delta_{32}(\gamma_{21}+\gamma_{42})+\Delta_{42}(\gamma_{21}+\gamma_{32})\Big)
\nonumber\\
             & & -\Delta^2_p\Big(\gamma_{21}+\gamma_{32}+\gamma_{42}\Big)
			           +\gamma_{21}\gamma_{32}\gamma_{42}
\nonumber\\
	           & & -\gamma_{21}\Delta_{32}\Delta_{42}+\gamma_{42}\big|\Omega_{32}\big|^2+\gamma_{32}\big|\Omega_{42}\big|^2  
\end{eqnarray}
and
\begin{eqnarray}
\label{I}	  
           B &=& \Delta_p\Big(\gamma_{21}\gamma_{32}+\gamma_{21}\gamma_{42}+\gamma_{32}\gamma_{42}
                 +\big|\Omega_{32}\big|^2+\big|\Omega_{42}\big|^2 \nonumber\\
             & & -\Delta_{32}\Delta_{42}\Big) -\Delta^2_p\Big(\Delta_{32}+\Delta_{42}\Big)
			           -\Delta^3_p+\Delta_{42}\big|\Omega_{32}\big|^2\nonumber\\	 
  	         & & +\Delta_{32}\big|\Omega_{42}\big|^2+\gamma_{21}\Big(\gamma_{42}\Delta_{32}+\gamma_{32}\Delta_{42}\Big).
\end{eqnarray}	
We then obtain $\chi'$ and $\chi''$ by separating the real and imaginary parts of equation (\ref{probe coherency}) and substituting in equation (\ref{susceptibility}) as
\begin{eqnarray}
\label{eq:chi_real}
     \chi' &=& \left(\frac{Nd^2_{21}}{\epsilon_0\hbar}\right)\frac{\Omega^*_p}{Z}\Bigg\{A\Big(\Delta_p(\gamma_{42}+\gamma_{32})+\gamma_{32}\Delta_{42}+\gamma_{42}\Delta_{32}\Big)
\nonumber\\    
           & & +B\Big((\Delta_{32}+\Delta_{p})(\Delta_{42}+\Delta_{p})-\gamma_{32}\gamma_{42}\Big)\Bigg\}
\end{eqnarray}
and
\begin{eqnarray}
\label{eq:chi_Image}
     \chi'' &=& \left(\frac{Nd^2_{21}}{\epsilon_0\hbar}\right)\frac{\Omega^*_p}{Z}\Bigg\{A\Big((\Delta_{32}+\Delta_{p})(\Delta_{42}+\Delta_{p})-\gamma_{32}\gamma_{42}\Big) \nonumber\\    
           & & -B\Big(\Delta_p(\gamma_{42}+\gamma_{32})+\gamma_{32}\Delta_{42}+\gamma_{42}\Delta_{32}\Big)\Bigg\}.
\end{eqnarray}
Here we have defined $Z=DD^*$, where $D=A+iB$. Below we shall suppose the coefficient $\frac{N|\vec{d}_{21}|^2}{\epsilon_0\hbar}$ to be unity. This setting will only affect the susceptibility. In other words, in what follows $\chi$ is displayed in units of $\frac{N|\vec{d}_{21}|^2}{\epsilon_0\hbar}$.
We will use Eqs.~(\ref{eq:chi_real}-\ref{eq:chi_Image}) to obtain the dispersion and absorption of the probe field as a function of the probe detuning. 
%
%
%
\subsection{Dressed state analysis}
\label{subsec:E}

The modification of the absorption and dispersion of the probe field can be understood by using the semi classical dressed state picture, since it provides useful insight into the origin of the interference mechanism \cite{Haroche}. For convenience, we take $\omega_2=0$.
Furthermore, hereafter we shall suppose that all decay rates are equal ($\gamma_{21}=\gamma_{32}=\gamma_{42}=\gamma$) and we shall express all frequencies in units of $\gamma$.
We define the rotated states:
\begin{equation}
\kett{\tilde 3}= e^{i\omega_{c1}t}\ket{3}\;,\;
\kett{\tilde 4}= e^{i\omega_{c2}t}\ket{4}\;,\;
\kett{\tilde 2}= \ket{2}\;,\;
\kett{\tilde 1}= \ket{1}~.
\end{equation}
The total Hamiltonian in the Schroedinger picture reads (here and in the following $\hbar=1$)
\begin{equation}
\begin{array}{lcl}
H&=&\omega_1\kett{1}
\braa{1} +
\omega_3\kett{3}\braa{3}
+\omega_4\kett{4}\braa{4}\\
&&-\Big(
\Omega_{32}e^{+i\omega_{c1}t}\kett{3}\braa{2}
+
\Omega_{42}e^{+i\omega_{c2}t}\kett{4}\braa{2}
+ \textrm{h.c.}\Big)\\
&=&\omega_1\kett{\tilde 1}
\braa{\tilde 1} +
\omega_3\kett{\tilde 3}\braa{\tilde 3}
+\omega_4\kett{\tilde 4}\braa{\tilde 4}\\
&&-\Big(
\Omega_{32}\kett{\tilde3}\braa{\tilde2}
+
\Omega_{42}\kett{\tilde4}\braa{\tilde2}
+ \textrm{h.c.}
\Big)~,
\end{array}
\end{equation}
where we have neglected the terms proportional to $\Omega_p$.
Finally, the Hamiltonian in the rotating frame is defined by the equation $i\frac{\partial \ket{\tilde Q}}{\partial t}=H^{rot}\ket{\tilde Q}$, where $Q=1,2,3,4$, and reads
%
%
\begin{equation}
\begin{array}{lcl}
\label{rotating H}     
     H^{rot} &=& \Delta_{42}\left|\tilde4\right\rangle\left\langle\tilde4\right|
           +\Delta_{32}\left|\tilde3\right\rangle\left\langle\tilde3\right|+\Delta_p \kett{\tilde1}\braa{\tilde1} \\ 
       & & -\Big(\Omega_{32}\left|\tilde3\right\rangle\left\langle\tilde2\right| 
           +\Omega_{42}\left|\tilde4\right\rangle\left\langle\tilde2\right|+h.c.\Big)~,
\end{array}
\end{equation}
where we have neglected, once again, the terms proportional to $\Omega_p$.
The diagonalization of the hamiltonian (\ref{rotating H}) leads to the dressed states, which are schematically shown in figure \ref{fig:four-level-system}(b). Being the ground state not coupled to any other state, we may just solve the corresponding characteristic equation for the three other eigenstates, which is:
\begin{eqnarray}  
\label{characteristic_eq}    
         & &  \lambda\Big(\lambda^2-\lambda(\Delta_{32}+\Delta_{42})
              +\Delta_{32}\Delta_{42}-(\Omega^2_{32}+\Omega^2_{42})\Big)
\nonumber\\ 
         & &  +(\Omega^2_{32}\Delta_{32}+\Omega^2_{42}\Delta_{42})=0,
\end{eqnarray} 
where we supposed $\Omega_{32}$ and $\Omega_{42}$ to be real, without restriction of generality.
In general, this equation can be solved analytically. However, for simplicity, below we shall find the eigenstates (i.e., the dressed states) and the corresponding eigenvalues separately for each case of interest in this work. Furthermore, for convenience we shall relabel $\kett{\tilde 1}\to\kett{1}$, $\kett{\tilde 2}\to\kett{2}$, $\kett{\tilde 3}\to\kett{3}$, $\kett{\tilde 4}\to\kett{4}$.

%
\subsubsection{Case 1}

We assume that both coupling fields are in resonance, $\Delta_{42}=\Delta_{32}=0$, and have the same strength, $\Omega_{32}=\Omega_{42}=\Omega$.
In this case, the eigenvalues $\lambda$ can be written as 
\begin{eqnarray}
\label{eigenvalues1}      
       \lambda_0=0~,\; \lambda_{\pm}=\pm\Omega\sqrt{2}~, 
\end{eqnarray}  
while the normalized eigenstates are given by
\begin{equation}
\begin{array}{lcl}  
\label{eigenstates1}   
      \ketm{0}&=&\ds\frac{1}{\sqrt{2}}\big(\ketm{3}-\ketm{4}\big)~,\\[0.3cm]
			\ketm{\pm}&=&\ds \frac{1}{\sqrt{2}}\left(
			\ketm{2} \mp \frac{1}{\sqrt{2}}\big(\ketm{3}+\ketm{4}\big)
			\right)~.
\end{array}
\end{equation}
%

%
\subsubsection{Case2}

We assume that $\Delta_{32}=0$, $\Delta_{42}=5$, and that the strengths of the two coupling fields are $\Omega_{32}=\Omega_{42}=3$. For this case, the eigenvalues $\lambda$ can be found to be 
\begin{eqnarray}
\label{eigenvalues2}      
       \lambda_0\simeq1.89~,\; \lambda_-\simeq-3.57~, \; \lambda_{+}\simeq6.69~,
\end{eqnarray} 
while the normalized eigenstates are given by
\begin{equation}
\begin{array}{lcl}  
\label{eigenstates2}   
      \ketm{0}&\simeq& -0.47\ketm{2}-0.46\ketm{3}+0.75\ketm{4} ~, \\[0.3cm]
			\ketm{+}&\simeq&0.48\ketm{2}-0.85\ketm{3}-0.21\ketm{4}  ~,\\[0.3cm]
			\ketm{-}&\simeq&-0.74\ketm{2}-0.26\ketm{3}-0.62\ketm{4}~.
\end{array}
\end{equation}
%
%
\subsubsection{Case 3}

We assume that the coupling fields have the same strength, $\Omega_{32}=\Omega_{42}=\Omega$, and opposite detuning $\Delta_{42}=-\Delta_{32}=\Delta$. The eigenvalues $\lambda$ are
\begin{eqnarray}
\label{eigenvalues3}      
       \lambda_0=0~,\; \lambda_{\pm}=\pm\mu~, 
\end{eqnarray}  
with $\mu=\sqrt{\Delta^2+2\Omega^2}$.
The eigenstates can be then written as
\begin{eqnarray}   
\label{eigenstates3}   
       \ketm{0}& =& N_{0}\left(
			\ketm{2} + \frac{\Omega}{\Delta}\left(
			\ketm{3}-\ketm{4}
			\right)
      \right)~,\nonumber \\			
\ketm{\pm} &=& N_{\pm}\left(
			\ketm{2}+\frac{\Omega}{\Delta\mp\mu}\ketm{3}\nonumber\right.\\
			&&\;\left.\mp\left(
			+\frac{\mu}{\Omega}+\frac{\Omega}{\Delta-\mu}
			\right)\ketm{4}
\right)~,
\end{eqnarray} 
where $N_0$ and $N_{\pm}$ are normalization constants.
%
%
\subsubsection{Case 4}
We assume $\Omega_{32}$=3, $\Omega_{42}$=6, $\Delta_{42}$=$-\Delta_{32}$=5.
The eigenvalues can be found to be 
\begin{eqnarray}
\label{eigenvalues4}      
       \lambda_0\simeq-2.05~,\; \lambda_+\simeq9.20~, \; \lambda_{-}\simeq-7.15~,
\end{eqnarray} 
while the normalized eigenstates are given by
\begin{equation}
\begin{array}{lcl}  
\label{eigenstates4}   
      \ketm{0}&\simeq& 0.60\ketm{2}-0.61\ketm{3}+0.51\ketm{4} ~, \\[0.3cm]
			\ketm{+}&\simeq&0.57\ketm{2}-0.12\ketm{3}-0.81\ketm{4}  ~,\\[0.3cm]
			\ketm{-}&\simeq&-0.56\ketm{2}-0.78\ketm{3}-0.28\ketm{4}~.
\end{array}
\end{equation}
%
%
%
%
%
\section{Results and discussion}
\label{Sec:III}

\begin{figure}[b]
\centering\includegraphics [scale=0.65]{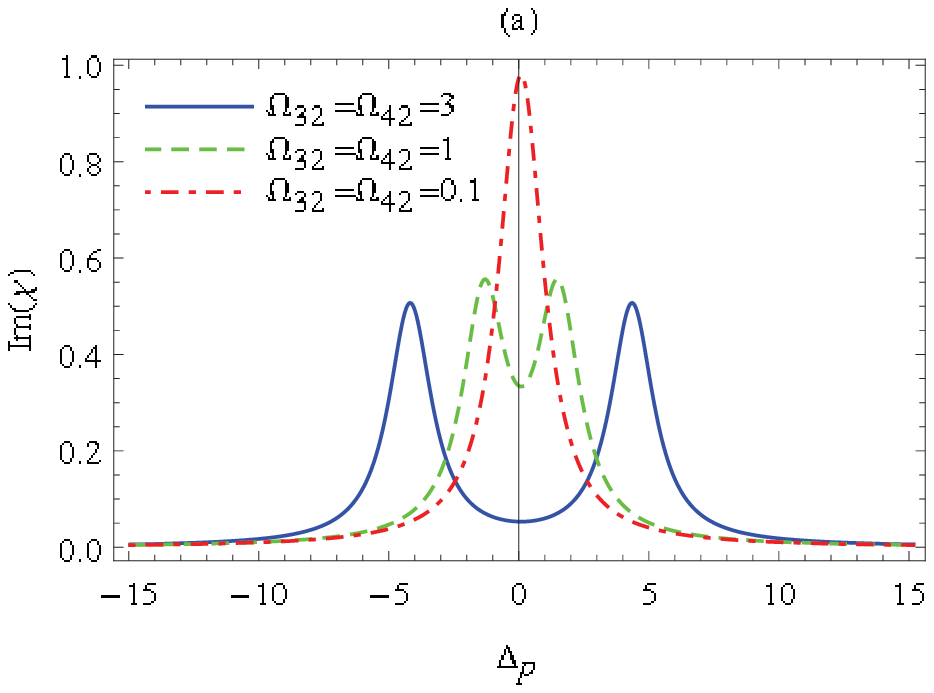}
\centering\includegraphics [scale=0.65]{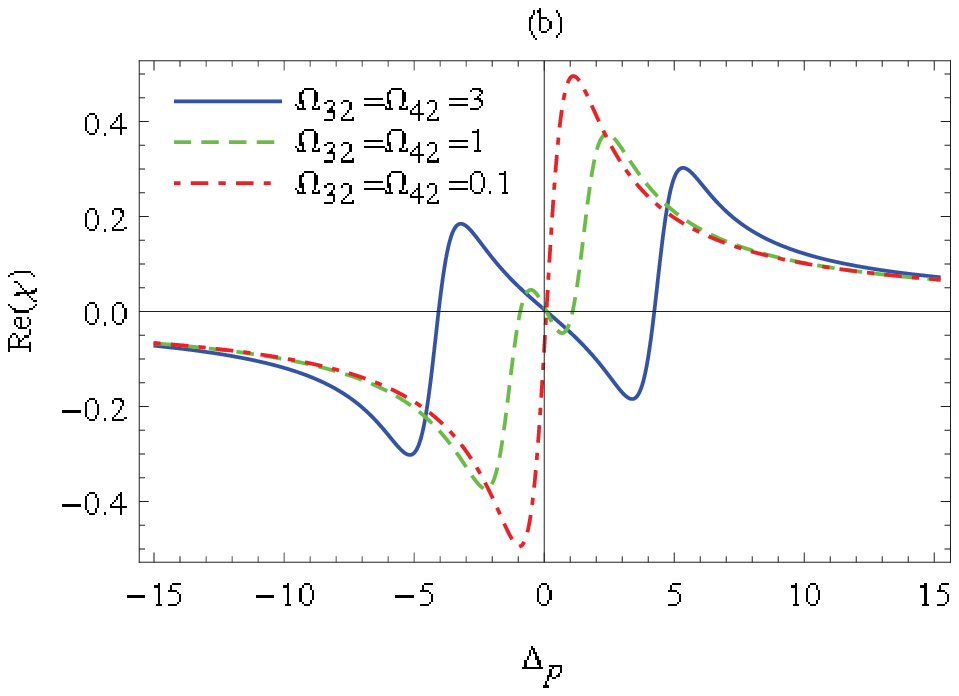}
\caption{Absorption (a) and dispersion (b) of the probe field for different coupling field strengths. 
Parameters are set as: $\Omega_p = 0.01$, $\Delta_{42}=\Delta_{32}=0$. 
Re($\chi$) and Im($\chi$) are in units of $\frac{N|\vec{d}_{21}|^2}{\epsilon_0\hbar}$, while $\Delta_p$ is in units of $\gamma$.
\label{figure2}}
\end{figure}
\begin{figure}[b]
\centering\includegraphics [scale=0.65]{{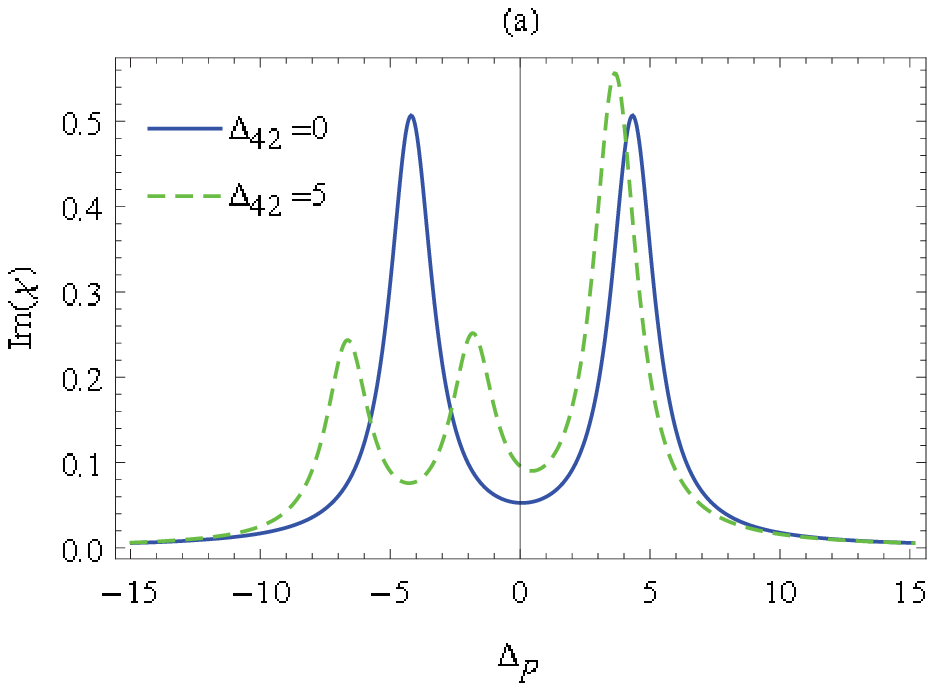}}
\centering\includegraphics [scale=0.65]{{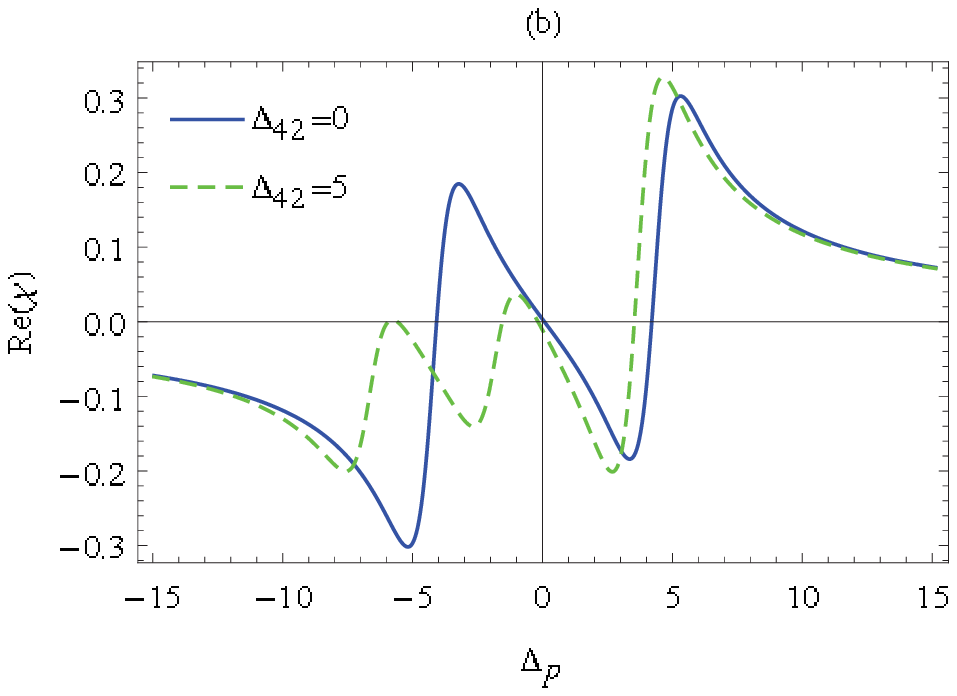}}
\caption{Absorption (a) and dispersion (b) of the probe field for different detunings $\Delta_{42}$. Parameters are set as: $\Omega_{p}=0.01$, $\Delta_{32}=0$, $\Omega_{42}=\Omega_{32}=3$. 
Units are the same as in Fig. \ref{figure2}.
\label{Fig4}}
\end{figure}
Below we shall study the absorption and the dispersion of the probe field by analyzing the susceptibility,  $\chi$. 
%
%
We set $\Omega_p=0.01$.
Results will be analyzed according to the dressed state analysis conducted in Sec. \ref{subsec:E}.

Figure~\ref{figure2} displays the absorption (a) and the dispersion (b) of the probe field as a function of its detuning $\Delta_p$, if we suppose the coupling fields to have the same strength and to be in resonance: $\Omega_{42}=\Omega_{32}$, $\Delta_{42}=\Delta_{32}=0$. As seen from the figure, if the strength of the coupling fields is very weak (red-dashed-dot curve), the probe field is absorbed at resonance and the dispersion shows an anomalous behavior while crossing the resonance. If the strength of the coupling field is increased to $\Omega_{42}=\Omega_{32}=3$, the absorption shows an EIT window with two well-separated equally-strong peaks at $\Delta_p=\pm 3\sqrt{2}$. These results can be easily understood from equations \eqref{eigenvalues1}-\eqref{eigenstates1}. Recalling that the initial state $\ketm{1}$ can couple only with $\ketm{2}$ due to parity selection rules, it follows that $\ketm{1}$ cannot be coupled with $\ketm{0}$, since this latter does not possess any admixture (component) of $\ketm{2}$. Therefore, the absorption peak related to the transition $\ket{1}\to\ket{0}$ is missing. However, $\ketm{1}$ can couple evenly with states $\ket{\pm}$, as they both possess the same component of $\ket{2}$. The two equally-strong peaks in figure \ref{figure2}(a) are therefore associated with the transitions $\ket{1}\to\ket{\pm}$. As for dispersion, its slope changes from positive (superluminal light) to negative (subluminal light) in the probe field resonance, due to the splitting of the absorption peak. This phenomenon has been first studied in a coherent-driven three level atom \cite{Boller:91}.

It is important to be able to control EIT for application in optical communications and quantum information theory. In order to control EIT here, we consider how the change of the detuning $\Delta_{42}$ affects the EIT window if we keep the coupling field $\Omega_{32}$ in resonance ($\Delta_{32}=0$) and, as in the previous case, if we set the coupling fields to have the same strength, $\Omega_{42}=\Omega_{32}=3$. Figure~\ref{Fig4} displays the absorption (a) and dispersion (b) of the probe field as a function of $\Delta_p$. If $\Delta_{42}=0$, the blue curve obtained in figure \ref{figure2} is recovered. However, if we increase the detuning of the coupling field $\Omega_{42}$ up to $\Delta_{42}=5$, the absorption peak at $\Delta_p= -3\sqrt{2}$ splits into two smaller peaks with another EIT window in the middle of them. These results can be once again understood from the dressed state picture. From equations \eqref{eigenvalues2}-\eqref{eigenstates2}, which represent the case under consideration, we see that all dressed states $\ket{0}$, $\ket{-}$, $\ket{+}$ possess some component of $\ket{2}$, so that all of them can be coupled to the initial state $\ket{1}$. For this reason, three absorption peaks are now present. Moreover, we can also identify each peak in figure \ref{Fig4}(a) as follows. As seen from equation\bl{s} \bl{\eqref{eigenvalues1}-}\eqref{eigenvalues2}, increasing the detuning $\Delta_{42}$ from $0$ to $5$ corresponds to shifting the energy of the state $\ket{0}$ upwards, from $0$ to $\simeq1.89$. This implies that the resonant peak for the absorption $\ket{1}\to\ket{0}$ (now possible) is obtained for values of the probe frequency larger than $\omega_{21}$ or, precisely, for the value $\Delta_p=\omega_{21}-\omega_p\simeq -1.89$. With the same reasoning, one can assign the peaks at $\Delta_p\simeq-6.69$, $3.57$ to the transitions $\ket{1}\to\ket{+}$, $\ket{-}$, respectively. Finally, we notice that the more admixture of $\ket{2}$ the dressed state possesses, the more pronounced the absorption peak is, as expected.
As for dispersion, the two EIT windows at $\Delta_p\simeq0$ and at $\Delta_p\simeq-3\sqrt{2}$ are both characterized by subluminal light propagation.
\begin{figure}[t]
\centering\includegraphics [scale=0.35]{{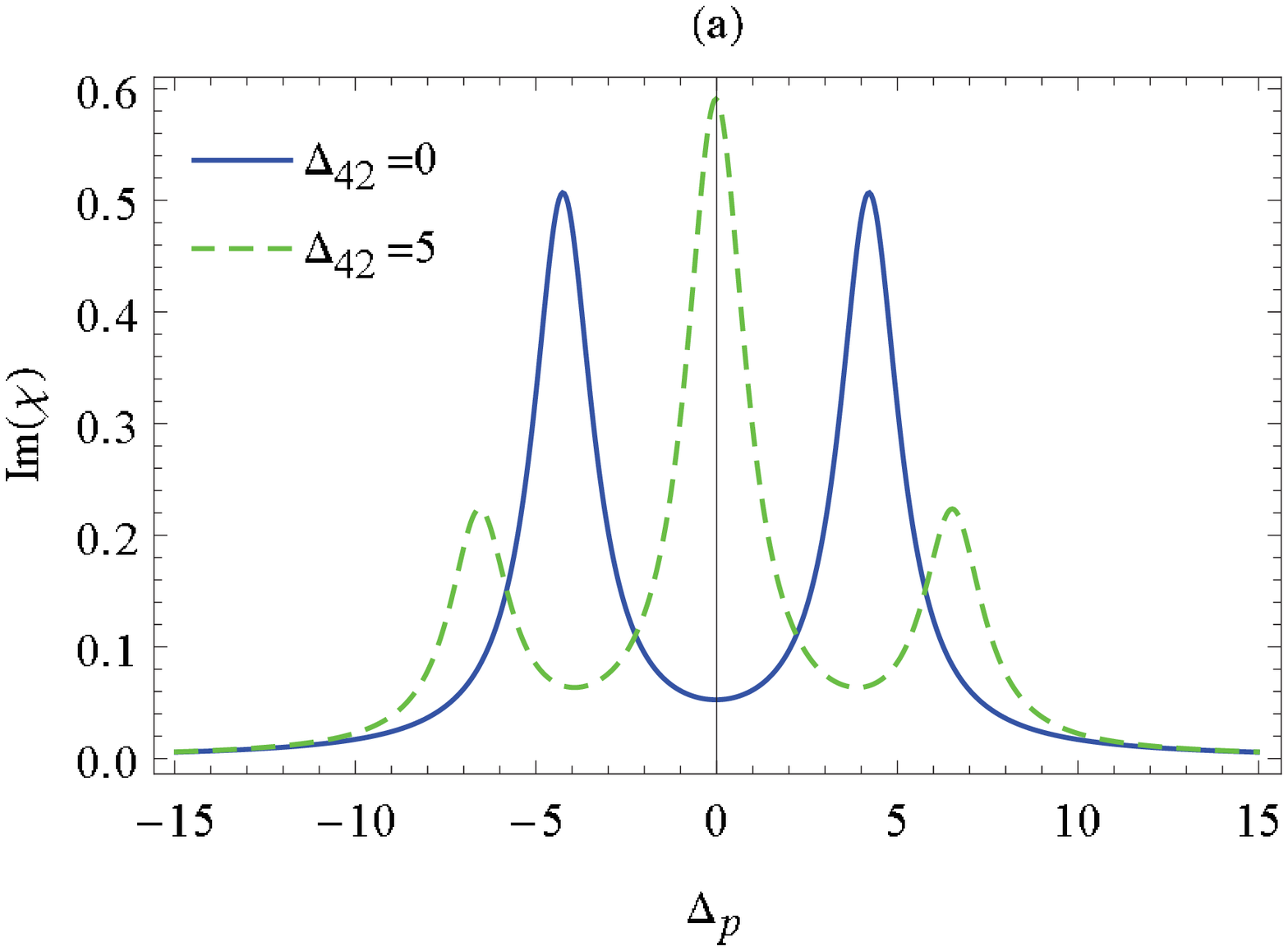}}
\centering\includegraphics [scale=0.65]{{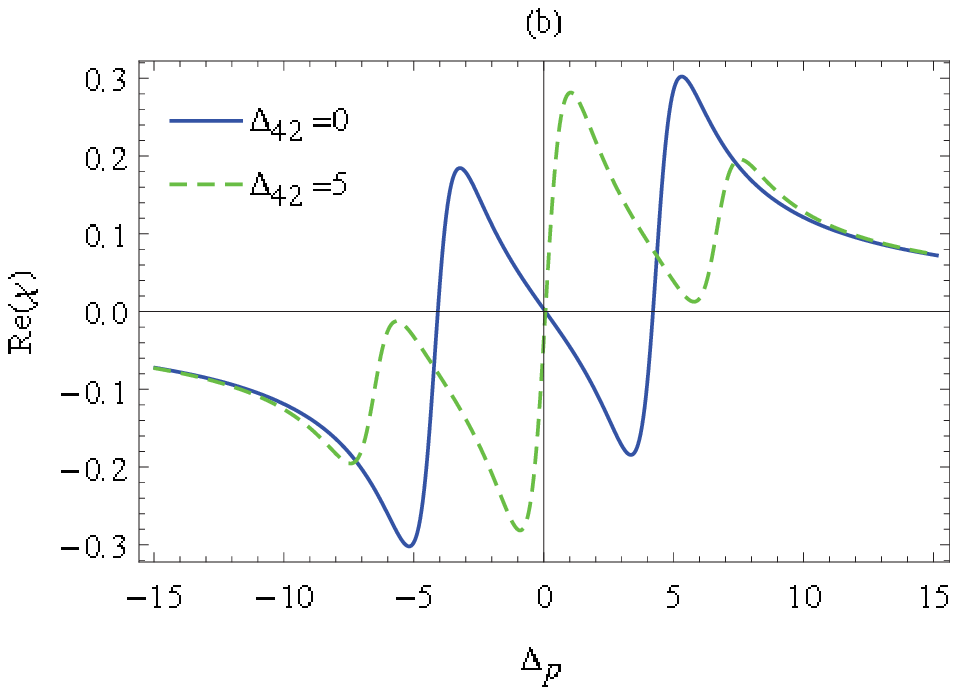}}
\caption{Absorption (a) and dispersion (b) of the probe field for different detunings $\Delta_{42}$. 
Parameters are set as:  $\Omega_{p}=0.01$, $\Omega_{42}=\Omega_{32}=3$, $\Delta_{32}=-\Delta_{42}$.
Units are the same as in Fig. \ref{figure2}.
\label{Fig5}}
\end{figure}

Another way to control EIT is to change the detuning of both coupling fields. Figure~\ref{Fig5} displays the absorption (a) and dispersion (b) of the probe field as a function of $\Delta_p$ if we set opposite detunings ($\Delta_{42}=-\Delta_{32}$) and, as in the previous cases, if we set the coupling fields to have the same strength ($\Omega_{42}=\Omega_{32}=3$). For the case $\Delta_{42}=0$, the blue curve obtained in figure \ref{figure2} is recovered. However, if we increase the control field detuning up to $\Delta_{42}=5$, light transparency switches to light absorption (as well as subluminal to superluminal light propagation) in the probe resonance. Although increasing $\Delta_{42}$ destroys the usual EIT window in the probe resonance, it leads to other two EIT windows which are located at $\Delta_p=\pm 3\sqrt{2}$. In these two windows, we have subluminal light propagation as evident from figure~\ref{Fig5}(b). In order to analyze these results, we use equations \eqref{eigenvalues3}-\eqref{eigenstates3}, by setting $\Omega=3$ and $\Delta=5$. First, we notice that each dressed states in equation \eqref{eigenstates3} possesses some admixture of $\ket{2}$, so that three absorption peaks must be present. As noted in the previous case, the more admixture of $\ket{2}$ the dressed state possesses, the more pronounced the absorption peak must be. With this and other arguments analogous to the previous case, one can identify the peak at $\Delta_p=0$ with the transition $\ket{1}\to\ket{0}$, and the peaks at $\Delta_p=\pm\sqrt{43}$ with the transitions $\ket{1}\to\ket{\mp}$.
\begin{figure}[t]
\centering\includegraphics [scale=0.35]{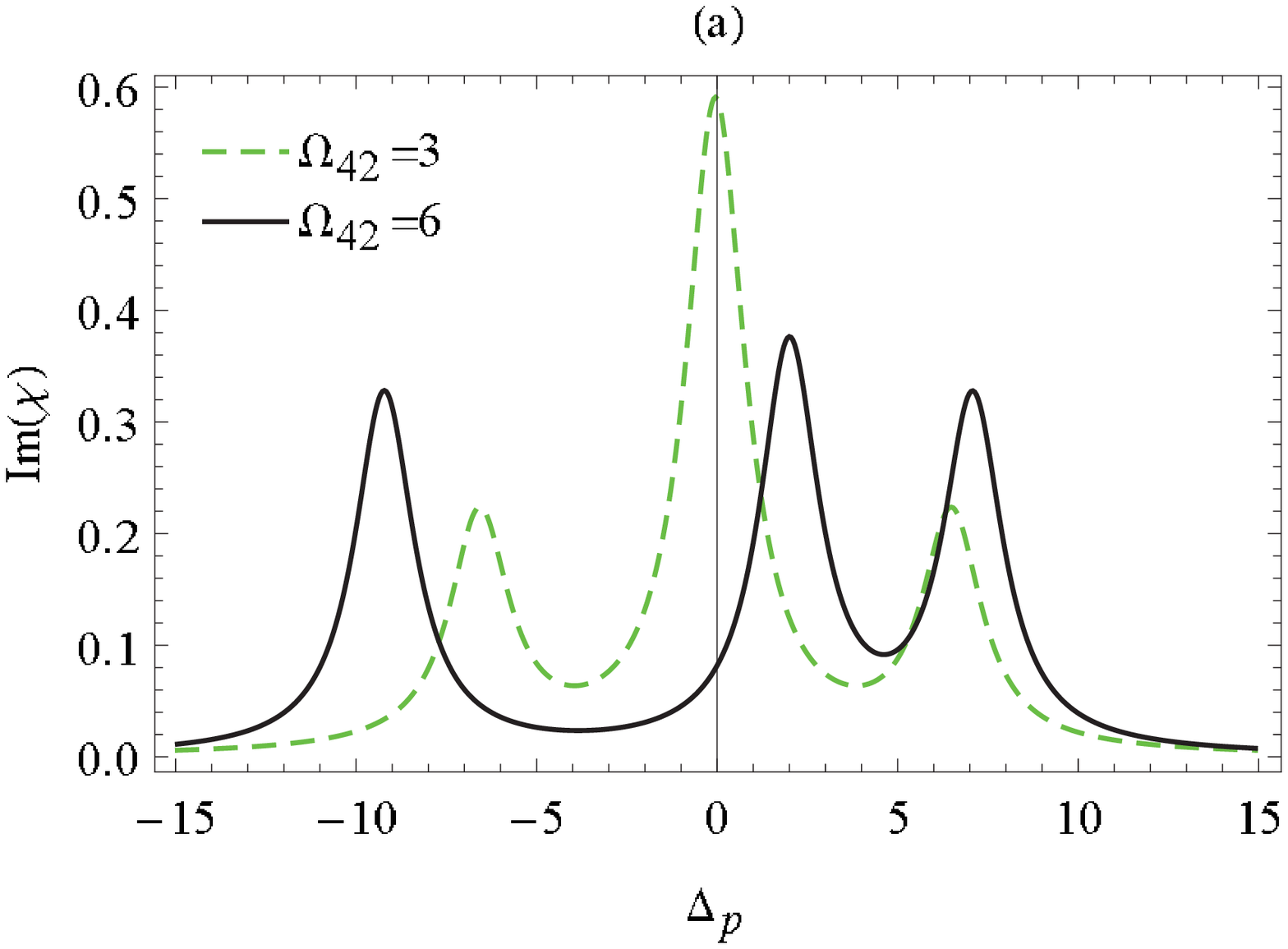}
\centering\includegraphics [scale=0.65]{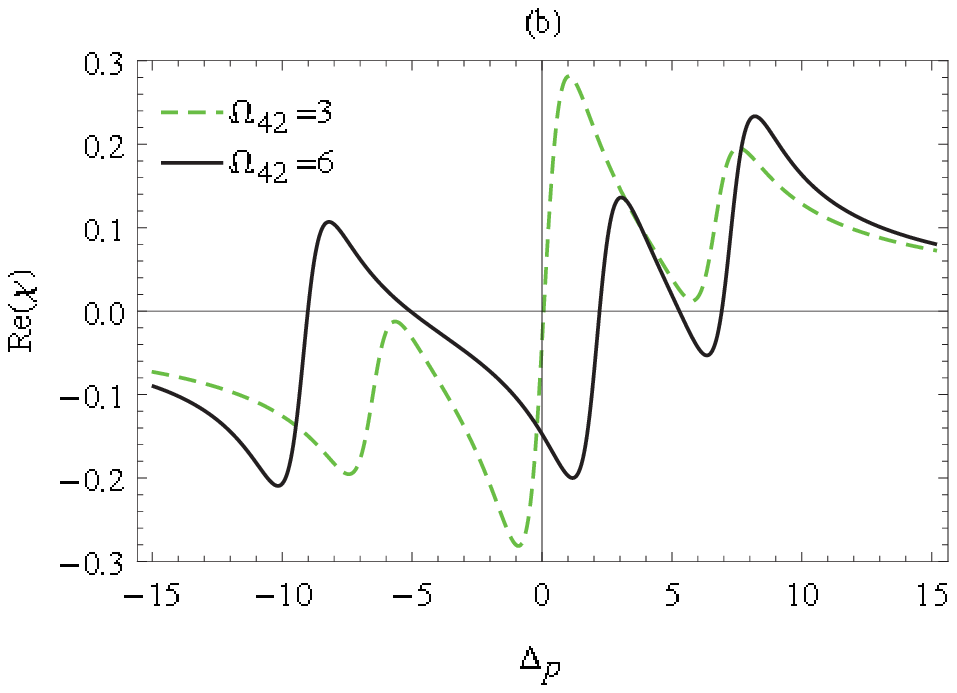}
\caption{Absorption (a) and dispersion (b) of the probe field for different coupling strength $\Omega_{42}$. 
Parameters are set as:  $\Omega_{p}=0.01$, $\Omega_{32}=3$, $\Delta_{42}=-\Delta_{32}=5$.
Units are the same as in Fig. \ref{figure2}.
\label{Fig6}}
\end{figure}

One of the desirable features in the EIT and group velocity control is intensity tunability. Our model shows that we can also control EIT windows by varying the intensity of coupling laser fields. Figure~\ref{Fig6} shows the absorption (a) and dispersion (b) of the probe field as a function of $\Delta_p$ for different intensities of the control field $\Omega_{42}$, if we set $\Omega_{32}=3$ and $\Delta_{42}=-\Delta_{32}=5$. For $\Omega_{42}=3$, the green-dashed curve in figure \ref{Fig5} is recovered. 
As seen from figure \ref{Fig6}(a), by increasing $\Omega_{42}$ up to $6$, the EIT window at $\Delta_p= +3\sqrt{2}$ becomes narrower with larger absorption, while the left side window at $\Delta_p=-3\sqrt{2}$ becomes broader with less absorption.
Once again, we make use of the dressed state picture to understand these results. From equations \eqref{eigenvalues4}-\eqref{eigenstates4}, by using the same arguments as for the previous cases, we can identify the absorption peak at $\Delta_p\simeq2.05$ with the transition $\ket{1}\to\ket{0}$, while the other two absorption peaks at $\Delta_p\simeq-9.20, 7.15$ with the transitions $\ket{1}\to\ket{+}, \ket{-}$ respectively. We moreover notice that the absorption peak at $\Delta_p\simeq2.05$ is the strongest, as a consequence of the fact that the dressed state $\ket{0}$ contains the largest admixture of $\ket{2}$, when compared with the others. As for dispersion, by increasing the intensity of coupling laser field $\Omega_{42}$, we have faster subluminal light propagation and more transparency in the left side EIT window, while slower subluminal light propagation and less transparency in the right side EIT window. 
%
%
%
%
%
%
%
\section{Conclusions}
\label{sec:conclusion}

We studied the steady state behavior of the absorption and dispersion of a weak tunable probe field in a Y-type atomic system driven by two strong coupling laser fields. We studied the modification of the absorptive and dispersive behavior of the system by varying the detuning and the intensity of coupling fields. By adjusting the detuning of the coupling fields, the absorption spectrum is strongly modified, leading to two EIT windows. In contrast, by adjusting the intensity of the coupling fields, the width and transparency of the EIT windows are modified. Moreover, we constructed the dressed state representation of the system so as to explain the position and the strength of each probe absorption peak in the cases analyzed in this work. Finally, as for dispersive behavior, the group velocity of a light pulse can be controlled from subluminal to superluminal by adjusting the intensity and the detuning of the coupling laser fields. 
%
%
%
%
%
\section{acknowledgment}

This work was supported by the Research Council for Natural Sciences and Engineering of the Academy of Finland. 
F.F. acknowledges support by Funda\c{c}\~ao de Amparo \`a Pesquisa do estado de Minas Gerais (FAPEMIG) and Conselho Nacional de Desenvolvimento Cient\'ifico e Tecnol\'ogico (CNPq).
L.S. wishes to thank Prof. Erkki Thuneberg and Dr. Johannes Niskanen for their useful comments.
%
%
%
%
%

\end{document}